\documentclass[preprint2]{aastex}    %% AASTeX

\usepackage{graphicx}
\usepackage{url,graphics,epsfig}
\usepackage{amssymb}

\begin{document}

\shorttitle{K-Shell Photoabsorption of  Atomic Oxygen}
\shortauthors{McLaughlin et al.}


\title{High Precision {\it K}-Shell Photoabsorption 
           Cross Sections for Atomic Oxygen: Experiment and Theory}


\author{B. M. McLaughlin\altaffilmark{1,2} }
\affil{$^1$Centre for Theoretical Atomic, Molecular and Optical Physics (CTAMOP)\\
	School of Mathematics and Physics, Queen's University Belfast\\
	Belfast BT7 1NN, Northern Ireland, UK\\
	$^2$Institute for Theoretical Atomic and Molecular Physics (ITAMP)\\
	Harvard Smithsonian Center for Astrophysics \\
	 MS-14, Cambridge, MA 02138, USA}
\email{b.mclaughlin@qub.ac.uk}

\author{C. P. Ballance\altaffilmark{3}}
\affil{$^3$Department of Physics, 206 Allison Laboratory\\
         Auburn University, Auburn, Alabama 36849-5311, USA}
\email{ballance@physics.auburn.edu}

\author{ K. P. Bowen\altaffilmark{4} and D. J. Gardenghi\altaffilmark{4}}
\affil{$^4$Department of Chemistry, University of Nevada\\
         Las Vegas, Nevada 89154-4003, USA}
\email{bowenk4@gmail.com, dgardenghi@gmail.com}

\author{W. C. Stolte\altaffilmark{4,5,6}}
\affil{$^4$Department of Chemistry, University of Nevada\\
         Las Vegas, Nevada 89154-4003, USA\\  
         $^5$Harry Reid Center for Environmental Studies, University of Nevada\\
                        Las Vegas, Nevada 89154-4009 USA\\
        $^6$Advanced Light Source, Lawrence Berkeley National Laboratory\\
                 Berkeley, California 94720, USA}
\email{wcstolte@lbl.gov}

\begin{abstract}
Photoabsorption of atomic oxygen in the energy region below the $\rm 1s^{-1}$ threshold 
 in x-ray spectroscopy from {\it Chandra} and {\it XMM-Newton} is observed in a 
 variety of x-ray binary spectra.  Photoabsorption cross sections 
 determined from an R-matrix method with pseudo-states (RMPS)  
 and new, high precision measurements 
 from the Advanced Light Source (ALS) are presented.
High-resolution spectroscopy with E/$\Delta$E $\approx$ 4,250 $\pm$ 400 was obtained for
photon energies from 520 eV to 555 eV at an energy resolution of 124 $\pm$  12 meV FWHM. 
{\it K}-shell photoabsorption cross-section measurements were made with a re-analysis of previous 
experimental data on atomic oxygen at the ALS.
Natural linewidths $\Gamma$ are extracted for the $\rm 1s^{-1}2s^22p^4 (^4P)np~^3P^{\circ}$  and 
$\rm 1s^{-1}2s^22p^4(^2P)np ~^3P^{\circ}$ Rydberg resonances series and compared with theoretical predictions.
Accurate cross sections and linewidths are obtained for applications in x-ray astronomy.
Excellent agreement between theory and the ALS measurements is shown which
will have profound implications for the modelling of x-ray spectra and spectral diagnostics.
\end{abstract}


\section{Introduction}
The photoionization process is one of the important radiative
feedback processes in astrophysics \citep*{Miyake2010,Stancil2010}.  
The increase in pressure caused by photoionization can trigger strong dynamic
effects, such as photoionization hydrodynamics.  The challenge 
in combining hydrodynamics with photoionization
lies in the difference in time scales between the two process.
Photoionization (PI) and photoabsorption (PA) processes play important
roles in many physical systems, including a broad range
of astrophysical objects as diverse as quasi-stellar objects (QSOs), the atmosphere of
hot stars, proto-planetary nebula, HII regions, novae and supernovae.
The {\it Chandra} and {\it XMM-Newton} satellites currently 
provides an abundance of x-ray spectra from 
astronomical objects; high-quality atomic data is needed to interpret  such spectra 
\citep*{McLaughlin2001,Mueller2010,McLaughlin2011,Brickhouse2010,Soleil2011,McLaughlin2013}. 
Theoretical studies, recently made on atomic carbon and it ions indicated high quality atomic data  are necessary to
accurately model {\it Chandra} observations in the x-ray spectrum of the blazar
Mkn 421\citep*{McLaughlin2010,Soleil2011}.

In the soft x-ray region (5-45 \AA), spectroscopy, including 
{\it K}-shell transitions for atomic elements such as, C, N, O, Ne, S and Si, in neutral, 
or low stages of ionization and {\it L}-shell 
transitions of Fe and Ni, are a valuable tool for probing the extreme 
environments in active galactic nuclei (AGN's), x-ray binary systems, 
cataclysmic variable stars (CV's) and Wolf-Rayet Stars 
\citep*{McLaughlin2001,Witthoeft2009,Brickhouse2010,Skinner2010,Mueller2010,Soleil2011,McLaughlin2011},
and the ISM \citep*{Garcia2011,Soleil2011}. 
Interstellar oxygen is found in both gas and dust phases, although the exact molecular 
form of the dust remains unknown.  Therefore an accurate understanding of the gas phase  
constituents is necessary in order to measure the residual molecular and solid-phase components \citep*{Kaastra2012}.  

In the x-ray community, Electron-Beam-Ion-Trap (EBIT) measurements (used for calibrating resonance energies), 
have been carried out for the inner-shell 1s $\rightarrow$ 2$\ell$ transitions in 
He-like and Li-like nitrogen ions \citep*{EBIT1999}, Li-like, Be-like, 
B-like and C-like oxygen ions \citep*{Schmidt2004,Gu2005}. 
In EBIT experiments, the spectrum is contaminated and 
blended with ions in multiple stages of ionization, 
making spectral interpretation fraught with difficulties. Cleaner, higher-resolution spectra are obtained at
synchrotron radiation facilities; ALS, BESSY II, SOLEIL, ASTRID II and Petra III. 
EBIT experiments have the advantage of the production of pure ground state populations of  atoms or ions, 
extremely difficult to make with merged beams methods, routinely used  at synchrotron radiation facilities. 

Photoionization and photoabsorption cross sections used for  the modelling of astrophysical
phenomena has traditionally been provided by theory,
as limited experimental data is available across a wide range of  wavelengths. 
Until recently, the bulk of theoretical work has not been tested thoroughly 
by experiment \citep*{McLaughlin2001,Mueller2010,Brickhouse2010,Soleil2011,McLaughlin2011}. 
For atomic oxygen the availability of x-ray data on this system provided 
the motivation to perform theoretical {\it K}-shell photoionization investigations. 

Inner-shell excitation processes occurring with the interaction of a photon
on the $\rm 1s^22s^22p^4~^3P$ ground-state of atomic oxygen 
 produces strong resonances observed in the corresponding 
 cross  section (c.f., Figure 2 and Table 1), through promotion of the  1s $\rightarrow$ $np$ electron via the processes;
$$
 h\nu + {\rm O (1s^22s^22p^4~^3P)}  \rightarrow  {\rm O ~ (1s2s^2 \,2p^4~[^{2,4}P]} np ~{\rm ^3P}^{\circ}) 
 $$
 giving competing decay routes namely,
 $$
{\rm  O^{+}~ (1s^22s^22p^3~[^4S^o, ^2D^o, ^2P^o]) + e^- ({\it k^2_{\ell}}),}
$$
and
 $$
{\rm  O^{+}~ (1s^22s^22p^2~[^3P, ^1D, ^1S])} n{^{\prime}}{\ell^{\prime}} + e^- ({\it k^2_{\ell}),}
$$
 which the present theoretical approach attempts to simulate, 
where, $n$ = 2 -- 6 (observed in the experiment),  and  $k^2_{\ell}$ is the outgoing 
energy of the continuum electron with angular momentum $\ell$.
{\it K}-shell photoionization contributes to the ionization balance in a
more complicated way than outer shell photoionization. 
{\it K}-shell photoionization when followed by Auger decay couples
three or more ionization stages instead of two in 
the usual equations of ionization equilibrium \citep*{Petrini1994,Petrini1997}.

Early theoretical photoionization (PI) cross section calculations for {\it K}-shell processes on this complex 
performed by Reilman and Manson \citep*{Reilman1979}
used the Hartree-Slater wavefunctions of Herman and Skillman \citep*{HS1963,Yeh1993} 
and the Dirac-Slater wavefunctions \citep*{Verner1993}. 
Photoionization cross sections determined from central field approximations although excellent 
for high photon energies  yield unreliable results near thresholds (see Figure 3 (a)), 
where resonance features dominate the cross sections \citep*{Yeh1993,Soleil2011}.

State-of-the-art  {\it ab initio} calculations for photoabsorption cross-sections and 
Auger inner-shell processes were first investigated 
on this system, using the standard R-matrix approach  \citep*{McLaughlin1998}, 
for modelling the resolved interstellar O~{\it K}, Ne~{\it K}, and Fe~{\it L}-edge absorption 
spectra in the {\it Chandra} x-ray Observatory Low-Energy Transmission Grating Spectrometer (LETGS)
 spectrum of the low-mass x-ray binary X0614+091 \citep*{Paerels2001}.
This work was extended using the optical potential technique  \citep*{McLaughlin2000} 
 to account for Auger broadening of resonances below the $\rm 1s^{-1}$ threshold 
and to analyze the high-resolution spectroscopy of the oxygen {\it K}-shell interstellar absorption edge in seven x-ray
binaries \citep*{Juett2004}. Garcia and co-workers \citep*{Garcia2005}, using the optical potential
method within the Breit-Pauli R-matrix formalism \citep*{Burke2011} extended this work to the oxygen iso-nuclear sequence 
and to investigate the x-ray absorption structure of atomic oxygen in the interstellar medium by analyzing {\it XMM-Newton}
observations of the low-mass x-ray binary Sco X-1 \citep*{Garcia2011}.

\section{Experiment}\label{secexp}
Cross sections for atomic  oxygen  {\it K}-shell photoionization were measured over the photon range 520 eV to 555 eV.
Our new results were obtained on undulator beamline 11.0.2 at the ALS, 
previous measurements were performed on bend-magnet beamline 6.3.2 \citep*{Stolte1997}. 
The present experimental measurements have covered the complete K-shell region, in a single scan, 
rather than constructing the spectra piece-meal like as was done previously  \citep*{Stolte1997}.
A resolution of 4250 $\pm$ 400  ($\approx$ 124 $\pm$ 12 meV) at 526.8 eV, and 
a photon flux of over $10^{11}$ photons/s was provided by BL 11.0.2 when using 
slit widths of 20 $\mu$m. The experimental apparatus has been discussed 
previously in detail  \citep*{Samson1988,Stolte1997,Stolte2008}.

\begin{figure}[!htb]
\begin{center}
\includegraphics[scale=1.5,width=8.0cm]{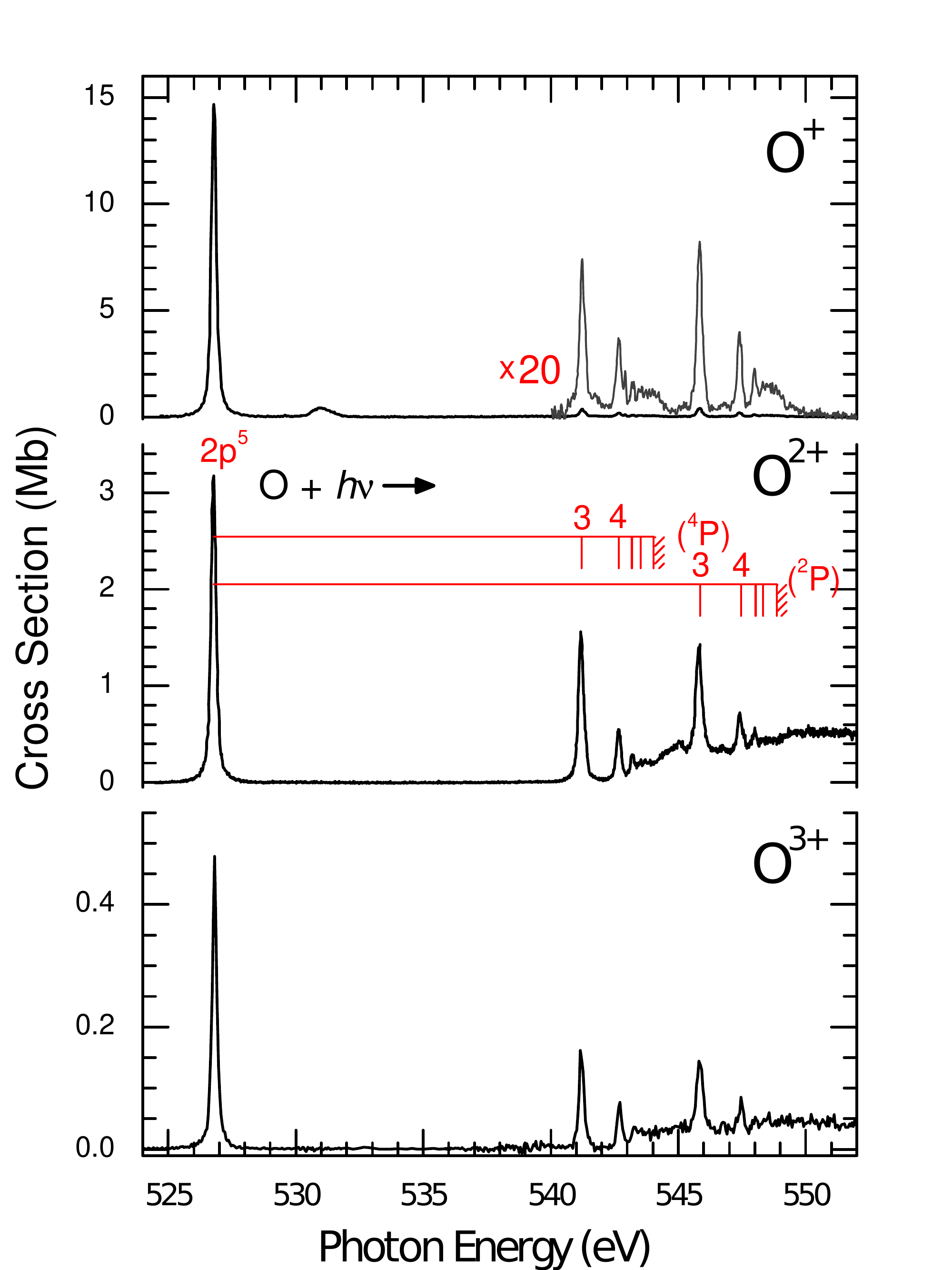}
\caption{\label{figexpt}(Colour online) Photoionization cross sections of O$^{+}$,  O$^{2+}$ and O$^{3+}$ produced
			                   by the decay of a $1s$ hole in atomic oxygen. The resonance
					lines represent the transitions  $\rm 1s2s^22p^5(^3P^{\circ})$, $\rm 1s2s^22p^4(^4P)np$
					and $\rm 1s2s^22p^4~(^2P)np$ with $n$ = 3 - 6. The current resolving power
					of the monochromator was 4250 $\pm$ 400  ($\approx$ 124 $\pm$ 12 meV) at a photon energy of 526 eV.}
\end{center}
\end{figure}

Similar to our earlier results \citep*{Stolte1997}, we used the Rydberg resonance features found in 
molecular oxygen near 541 eV to calibrate the photon energy scale, 
resulting in a maximum uncertainly of 40 meV.
During double-bunch operations of the ALS, a Wiley-McLaren style 
time-of-flight mass spectrometer \citep*{Wiley1955}, oriented with its axis parallel to the 
polarization vector of the incident synchrotron radiation, was used in conjunction with a 
microwave discharge system to determine the branching ratios for atomic oxygen at the
$\rm 1s2s^22p^5~(^3P^{\circ})$ resonance.

Both the microwave {\it on} (O and O$_2$ mixture) and {\it off} (O$_2$)
 independent spectral scans were divided by the incident photon flux.  The
molecular contribution to the microwave {\it on} spectrum was removed 
by a scaled subtraction of a microwave {\it off} spectrum at the 
O$_2$ : 1$s \rightarrow 1\pi_g^*$ resonance (530.5 eV).
Several residual peaks are created due to the molecular resonances 
being a few meV wider for the microwave {\it on} spectra.  We surmise this
difference is probably due to the larger thermal motion of the gas molecules with
the microwave discharge on.  The scans were finally corrected by removal of
the background produced by direct photoionization of the atomic oxygen valence shell. 
This was only necessary for O$^{+}$ or O$^{2+}$, being that O$^{3+}$ sat on a zero background. 
We performed two coarse photon energy scans, one for each ion, 
covering the range between 500 -- 600 eV. The O$^{+}$ or O$^{2+}$ 
signal produced by {\it K}-shell photoionization is superimposed on a 
nearly flat background caused by direct photoionization of the valence shell.
This background contributed 35\%  and 8\% to the total O$^{+}$ and O$^{2+}$ signals, 
respectively, just above the 1s$^{-1}$~$^2$P series limit. The independent ion 
specific photon energy scans could then be placed onto a relative scale 
by using the branching ratios (O$^{+}$/TIY = 80.07\%, O$^{2+}$/TIY = 17.32\%, 
and O$^{3+}$/TIY = 0.026\%, with TIY = total ion yield) measured 
with the time-of-flight mass analyzer on top of the $\rm 1s2s^22p^5~(^3P)$
resonance. Finally, the scans were placed onto an absolute scale (see Figure 1) by summing
their values above the $^2$P ionization limit and normalizing this sum to the difference of the
cross sections above and below the oxygen {\it K}-edge \citep*{Stolte1997,Stolte2008}.

\section{Theory}\label{subsecR-Matrix_Theory}
The $R$-matrix with pseudo-states method \citep*{Burke2011,codes,francis95, ballance06} was used to determine our 
theoretical results.  Cross section calculations were performed in $LS$-coupling 
retaining 910-levels (valence and hole-states) of the residual O$^{+}$ ion in the close-coupling expansion.
 Hartree-Fock $\rm 1s$, $\rm 2s$ and $\rm 2p$
\citep*{Clementi1974} and n=3 pseudo-orbitals of the O$^{+}$ residual ion 
  (included for core relaxation and correlations effects) were used,
  obtained by optimizing on the energy of the hole state; 
  $\overline{\rm 3s}$, $\overline{\rm 3p}$ and  $\overline{\rm 3d}$ on $\rm 1s2s^22p^4~^4P$ 
including the important configuration $\rm 1s^22s^22p^2\overline{\rm 3d}~^4P$
with the multi-configuration-Hartree-Fock (MCHF) 
atomic structure code \citep*{charlotte1991}.

For the O($\rm 1s^22s^22p^4$\, $\rm ^3$P) bound state 
we obtained 14.0344 eV for the ionization potential
using a triple electron promotion model, 
the NIST experimental value is 13.61806 eV, 
a discrepancy of  $\sim$ 3 \%, or  416 meV.  
A double electron promotion model gave 13.82184 eV, 
a discrepancy of  $\sim$ 1.5 \%, or  204 meV, 
yielding closer agreement with the NIST tabulated value as the RMPS approach 
provides more highly correlated wave functions.
In previous work  \citep*{McLaughlin2000}  obtained 
an underestimate of the ionization potential $\sim$ 2.5 \%, or 338 meV
compared to experiment, due to limited correlation included. 

In  the collision calculations for atomic oxygen, twenty
continuum functions and a boundary radius of 7.27 Bohr radii was used. 
Two and three-electron promotion scattering models were investigated giving similar results.
The collision problem was solved  using an energy grid of 
2$\times$10$^{-7}$ Rydbergs ($\approx$ 2.72 $\mu$eV) allowing
detailed resolution of resonance features  in the cross sections.  

The peaks  found in  the photoabsorption cross section spectrum were 
fitted to Fano profiles \citep*{Fano1968} instead of the energy derivative 
of the eigenphase sum technique \citep*{keith1998,keith1999}. 
Theoretical values for the natural linewidths $\Gamma$ (meV) are presented in Table 1 
and compared with current and previous 
ALS measurements \citep*{Stolte1997} and with prior investigations.

\begin{figure}[tb]
\begin{center}
\includegraphics[scale=2.0,width=8.5cm]{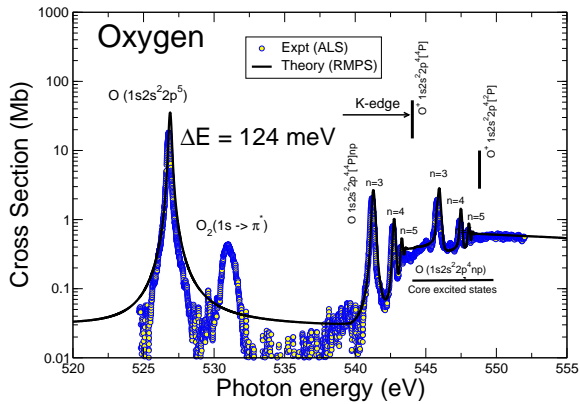}
\caption{\label{fig124meV} (Colour online) Atomic oxygen photoabsorption cross sections 
								taken at 124 meV FWHM  compared with theoretical estimates. 
								The R-matrix calculations shown are  from the
								RMPS method (solid black line, present results)
								convoluted with a Gaussian
								 profile of 124 meV FWMH.
								Table 1 designates the resonances and their properties.}
\end{center}
\end{figure}

\section{Results and Discussion}\label{secResults}
In the photon energy range (520 eV --  555 eV) explored, intense structure 
is observed in the cross section between 520 eV and 530 eV, from 
the strong $\rm 1s \rightarrow 2p$ transition in the atomic oxygen spectrum. 
Figure 2 shows our present experimental and theoretical results 
for the photon energy range of 520 eV -- 555 eV illustrating all the 
additional $\rm 1s \rightarrow np$ transitions  (n $\ge$ 3) in the spectrum.  
Previous experimental measurements \citep*{Stolte1997} were 
actually measured at a photon energy resolution of 
135 meV and not 182 meV, current ALS measurements are at 124 meV FWHM.  
Convolution of the RMPS theoretical results with a Gaussian function of 124 meV FWHM 
was used to compare directly with the ALS measurements.
Resonances observed in the experimental measurements  were fitted with Voigt profiles 
to determine the natural linewidths using a Gaussian function of 124 meV FWHM for each peak.
The photon energy was calibrated to an energy uncertainty of approximately $\pm$ 40 meV. 

Previous measurements of Stolte and co-workers \citep*{Stolte1997} were re-analyzed 
 with the present results.  The measured spectra for O$^+$, O$^{2+}$ 
 and O$^{3+}$ production were fitted with the program 
WinXAS$^\copyright$ of Thorsten Ressler, Hamburg, Germany  \citep*{Ressler1998}
and its near edge x-ray absorption fitting routines. 
Due to an incomplete data set,  previous measurements for O$^{3+}$ \citep*{Stolte1997} were not fitted. 
In previous measurements \citep*{Stolte1997}, a width of 231 meV 
was  cited with a resolution of 182 meV.  On refitting the previous results \citep*{Stolte1997} it 
was discovered that it was not possible to arrive at a proper fit with a resolution of  3000.  
Current multi-function fits, using Voigt and arctan functions, 
determined the resolution to be 3800 $\pm$ 150 ($\approx$ 135 $\pm$ 5 meV).  
The 1s2s$^2$2p$^4$($^4$P$^e$)6p and 1s2s$^2$2p$^4$($^2$P$^e$)6p states 
cannot be properly fitted, since they are completely hidden. 

Rydberg's formula was used to determine the resonance energies, given by, 
 \begin{equation}\label{rydberg}
\epsilon_n  =  \epsilon_{\infty} -  \frac{{\cal~Z}^2} { \nu^{2}}.   
\end{equation} 
\noindent
Where, $\epsilon_n$ is the resonance transition energy, in Rydbergs,
 $\epsilon_{\infty}$ the ionization potential and the resonance series 
 limit.  The principal quantum number $n$, 
the effective quantum number $\nu$ and the quantum defect $\mu$ 
are related by $\nu$ = $n - \mu$ \citep*{Seaton1983,Hinojosa2012}.
Converting all quantities to eV, members of the 
Rydberg series are represented by,
\begin{equation}\label{eV}
 E_n = E_{\infty} - \frac{{\cal{Z}}^2 {\cal R}}{(n - \mu)^2} .
\end{equation} 
\noindent
$E_n$ is the resonance energy position,  $E_{\infty}$ the ionisation limit, 
$\cal{Z}$ is the charge of the core (in this case, ${\cal Z}$ = 1) and ${\cal R}$ is 
13.6057 eV \citep*{Seaton1983,Hinojosa2012}. 
 
 In Figure 2  we present the experimental cross section measurements from the ALS taken at
  4250 $\pm$ 400  ($\approx$ 124 $\pm$ 12 meV) resolution 
  compared to our theoretical work. In the non-resonant region, above the K edge, at 550 eV, 
  theory gives a value of 0.560 Mb, the ALS experimental 
  measured value is 0.559 Mb, a discrepancy of  0.03\%.
 Figure 2 shows the excellent agreement between theory and experiment 
over the entire energy region and Figure 3 (a) illustrates the RMPS results with 
the central field approximation \citep*{Yeh1993} results.  Strong resonance 
features near the K-edge present in the  state-of-the-art RMPS cross sections are 
absent from the central field calculations.  In Figure 3(b) un-convoluted, RMPS 
cross sections with the optical potential R-matrix results \citep*{McLaughlin2000} are presented. 
Note, the discontinuity ($\sim$ 538 eV) in the optical potential 
R-matrix cross section results absent from the RMPS calculations.
Finally, Figure 4  compares quantum defects $\mu$ obtained 
from the RMPS  and the optical potential 
methods \citep*{McLaughlin2000,Garcia2005} with experiment. 
  
 Table 1 presents our experimental and theoretical results, 
 for resonance energies, resonance strengths $\overline{\sigma}_{n\ell} ({\rm Mb~ eV})$, 
 quantum defects $\mu$ and natural linewidths $\Gamma (\rm meV)$ with previous theoretical 
 and experimental work \citep*{Stolte1997,Petrini1994, Menzel1996,Saha1994,Krause1994}.
 Table 1 includes the experimental values for the lifetime $\tau$ expressed in femto-seconds (fs), 
 determined via the uncertainty principle ($\Delta E \Delta t$ = $\hbar$/2).
  
\section{Conclusions}\label{secConclusions}
{\it K}-shell photoabsorption of atomic oxygen was investigated using the 
R-matrix with pseudo-states (RMPS) method along 
 with current and previous high resolution experimental measurements  made at the ALS.  
 Resonance features observed in the cross sections for photon energies in the range 520 eV -- 555 eV,
are identified as 1s $\rightarrow$ np transitions that are analyzed, natural 
linewidths $\Gamma$ (meV) extracted and lifetimes determined via the uncertainty principle. 
Excellent agreement (see Table 1) of  theoretical estimates, for the resonance parameters 
(resonance energies, natural line widths and quantum defects)
compared to experimental measurements is obtained with calculations for the
ionization potential accurate to within 1.5 \% of experiment.
We have delineated {\it all} of the resonances properties and made a 
 detailed comparison of experiment with current and previous theoretical investigations. 
Earlier theoretical work \citep*{McLaughlin2000,Garcia2005}  made a limited comparison 
 with experiment, for the quantum defects of the two resonance series with 
only the Auger width of $\rm 1s2s^22p^5~^3P^{\circ}$ state determined.
The present ALS high resolution (124 meV FWHM) of all the resonance peaks observed in the cross sections
over the {\it entire} energy range investigated allowed a direct comparison 
with state-of-the-art R-matrix calculations to be made. The only limitation of the present theoretical model 
is apparent from  Table 1 concerning the resonance strengths which might be addressed in the future
with an extended pseudo-states basis within a Breit-Pauli approximation.
Finally, the data is suitable to be incorporated into the astrophysical modelling codes CLOUDY
 \citep*{Ferland2003}, XSTAR \citep*{Kallman2001} and AtomDB \citep*{Foster2012}.
 
\begin{figure}[tb]
\begin{center}
\includegraphics[scale=2.0,width=8.5cm]{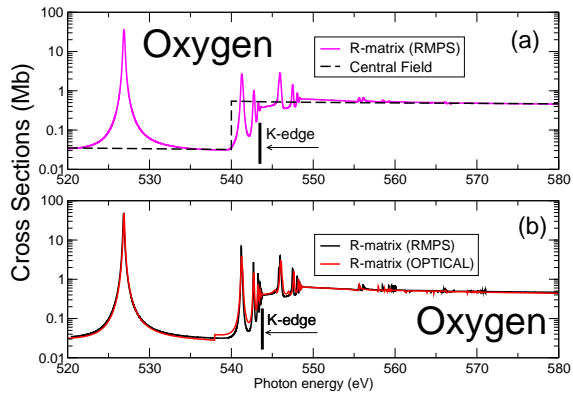}
\caption{\label{figasym} (Colour online)  (a) R-matrix (RMPS) cross sections (convoluted at 124 meV) compared  
								with central field approximation results \citep*{Yeh1993}, 
								(b) Cross sections (unconvoluted) from
								the RMPS (solid black line) and Optical Potential R-matrix calculations (solid red line).}
\end{center}
\end{figure}

\begin{figure}[tb]
\begin{center}
\includegraphics[scale=2.0,width=8.5cm]{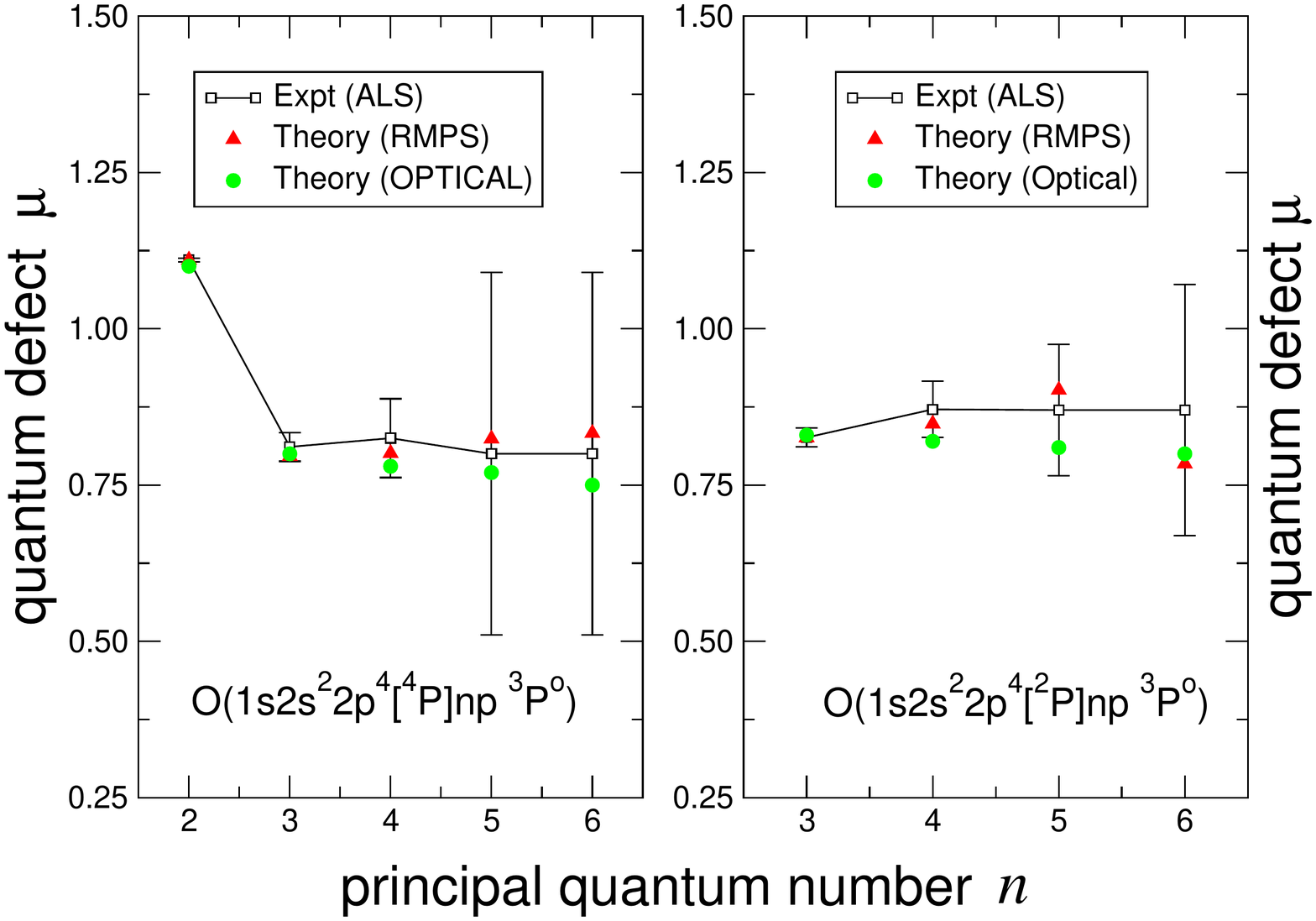}
\caption{\label{figqdef} (Colour online)  ALS experimental quantum defects with 
                                                                       those obtained from theory. The RMPS results, filled triangles, solid circles
                                                                       optical potential approach \citep*{McLaughlin2000,Garcia2005}.}
\end{center}
\end{figure}

The energy resolution on present day satellites,  {\it Chandra} and {\it XMM-Newton},  
is  $\sim$ 0.6 eV, a factor of 10 lower than available at current  ground based synchrotron 
radiation facilities, like the ALS, SOLEIL, ASTRID II, BESSY II or PETRA III, providing higher resolution and precision 
than obtained via satellites.  There are also issues concerning the 
calibration of spectra obtained from satellites as previously highlighted.  The ALS measurements 
have been calibrated to the resonance transition in molecular oxygen, which is known very accurately.
%
%+++++++++++++++++++++++++++++++++++++++++++++++++++++++++++++++++++++++++++++
%
%    Table 1 here
%
%    Here is an example of the general form of a table
%    Fill in the caption in the braces of the \caption{} command. Put the label
%    that you will use with \ref{} command in the braces of the \label{} command.
%    Insert the column specifiers (l, r, c, d, etc.) in the empty braces of the
%    \begin{tabular}{} command.
%
%+++++++++++++++++++++++++++++++++++++++++++++++++++++++++++++++++++++++++++++
%

\begin{deluxetable}{cccccccc}
\tabletypesize{\scriptsize}
%\rotate
\tablecaption{\scriptsize Theory and experiment values for the resonance parameters for the $\rm 1s2s^22p^5(^3P^o)$ and  $\rm 1s2s^22p^4(^{2,4}P)np~ ^3P^o$,  
	       Rydberg series ($n \ge 3$) converging to the $\rm 1s^{-1}~^4P$ and $\rm 1s^{-1}~^2P$ series limits of atomic oxygen.} 
\tablehead{
Resonances				& Energy 					& Energy 				&  $\mu$				    	& $\mu$				&  $\Gamma$ 		          & $\Gamma$   		         	          & Lifetime \\
   (Symmetry)             			&   (eV)                                     	&   (eV)               		&                     	 		    	&					&	(meV)			 & (meV)	                  	                   & $\tau$  (fs)    \\
   \\
   Label					& Theory					& Expt				&Theory					& Expt				& Theory				 & Expt					& Expt \\
($\rm ^3P^{\circ}$ ) 			& (R-matrix$^a$)			& (ALS$^b$)			& (R-matrix$^a$)	 	   	& (ALS$^b$)			&  (R-matrix$^a$)	          & (ALS$^b$)    		                   &(ALS$^b$)    }
\startdata
$\rm 2p^5$				& 526.83 				   	& 526.79 $\pm$ 0.04	& 1.110					& 1.110 $\pm$ 0.003	& 150$^a$	 	          & 148 $\pm$ 11$^b$		          &2.22 $\pm$ 0.17\\
						&						&					&						&					& 169$^e$			& 153 $\pm$ 12$^c$			&-			\\
						&						&					&						&					& 139$^f$				& 160 $\pm$ 9 $^d$			&-			\\			
$\rm 3p$   				& 541.23					& 541.19	$\pm$ 0.04	& 0.796				   	& 0.811 $\pm$ 0.023	& 143		                   & 167 $\pm$ 11 	  		&1.97 $\pm$ 0.13\\
$\rm 4p$   				& 542.70 					& 542.68	$\pm$ 0.04	& 0.801				   	& 0.825 $\pm$ 0.063	&  97		 	                   & 125 $\pm$ 11		     	&2.63 $\pm$ 0.23\\
$\rm 5p$   				& 543.25 					& 543.23	$\pm$ 0.04	& 0.824				   	& (0.800 $\pm$ 0.29)	&  70			                   & 119 $\pm$ 10 	  		&2.77 $\pm$ 0.23\\
$\rm 6p$					& 543.52 					& 543.51	$\pm$ 0.04	& 0.833				   	& (0.800 $\pm$ 0.29)	&  80				         & 196 $\pm$ 12			&1.68 $\pm$ 0.10\\
     -						& -						& -					& -					  	& -					& -		 		         & -				  	          &-			    \\
     -						& -						& -					& -					  	& -					& -		 		         & -				  	          &-			    \\
$\rm 1s^{-1}~^4P$			& 544.03				    	& 544.03 	$\pm$ 0.04	&					  	& 					& 				         & 				 		 &			    \\
\\
$\rm 3p$  		 			& 545.97  					& 545.83	$\pm$ 0.04	& 0.826					& 0.870  $\pm$	 0.015	& 187		                   & 196 $\pm$ 10 	     	         & 1.68 $\pm$ 0.10\\
$\rm 4p$					& 547.48					& 547.45  $\pm$ 0.04	& 0.848					& 0.871 $\pm$	 0.045	& 127				& 128 $\pm$ 10 			& 2.57 $\pm$ 0.20\\
$\rm 5p$					& 548.04					& 548.04	$\pm$ 0.04	& 0.902					&(0.870 $\pm$	 0.105)	& 102			    	& 156 $\pm$ 10 			& 2.11 $\pm$ 0.14\\
$\rm 6p$    				& 548.35	 				& 548.33	$\pm$ 0.04	& 0.784					&(0.870 $\pm$	 0.201)	& 110		                   & 157 $\pm$ 13		  	& 2.10 $\pm$ 0.17\\
       -						& -						& -					& -						& -					& - 		           	         & -				 	         &-			     \\
     -						& -						& -					& -					  	& -					& -		 		         & -				  	         &-			    \\
$\rm 1s^{-1}~^2P$  			& 548.85 				         & 548.85	$\pm$ 0.04	&						& 					&	     			         & 					         &	                      \\
						&						&					&						&					&					&						&                             \\
						&						&					&     Strengths				&	Strengths			&					&						&                             \\
						&						&					&	Theory				&	Experiment		&					&						&                             \\
						&  Resonance				&					&	(R-matrix$^a$)			&	(ALS$^b$)		& 					&						&                   \\
						&						&					&						&					&					&						&                             \\
						&	Series				& 					&$\rm\overline{\sigma}_{n\ell}$&$\rm\overline{\sigma}_{n\ell}$	&				&						&                             \\
						&$\rm 1s2s^22p^4(^4P)n\ell~^3P^{\circ}$	& 			&     	(Mb eV)				&	(Mb eV)			&					&						&                             \\
						&$\rm\overline{\sigma}_{2p}$	&					&     10.22					&	5.76	$\pm$ 1.45	&					&						&                             \\
						&$\rm\overline{\sigma}_{3p}$	&					&       0.88					&	0.68	$\pm$ 0.17	&					&						&                             \\
						&$\rm \overline{\sigma}_{4p}$	&					&       0.30					&	0.25	$\pm$ 0.06	&					&						&                             \\
						&$\rm\overline{\sigma}_{5p}$	&					&       0.13					&	0.08	$\pm$ 0.02	&					&						&                             \\
						&$\rm \overline{\sigma}_{6p}$	&					&       0.09					&	 -				&					&						&                             \\
\\
						&$\rm 1s2s^22p^4(^2P)n\ell~^3P^{\circ}$	&			&						&					&					&						&                             \\
						&$\rm \overline{\sigma}_{3p}$	&					&	1.12					&	0.58	$\pm$ 0.15	&					&						&                             \\
						&$\rm \overline{\sigma}_{4p}$	&					&	0.44					&	0.17	$\pm$ 0.04	&					&						&                             \\
						&$\rm \overline{\sigma}_{5p}$	&					&	0.20					&	0.07	$\pm$ 0.02	&					&						&                             \\
						&$\rm \overline{\sigma}_{6p}$	&					&	0.16					&	 -				&					&						&                             \\
\\
\enddata
\tablerefs{
$^a$R-matrix (RMPS), $\rm 1s^{-1}~^{2,4}P$ series limit are the ALS measurements of Stolte and co-workers \citep*{Stolte1997};
$^b$Experiment, resolution of 4250 $\pm$ 400 (124 $\pm$ 12 meV), to fit the current ALS data;
$^c$Experiment, resolution of 3800 $\pm$ 150 (135 $\pm$ 5 meV), to refit the previous ALS  data  \citep*{Stolte1997} ;
$^d$Experiment, Wisconscin, Synchrotron Radiation Center (SRC), Krause and co-workers \citep*{Krause1994,Menzel1996};
$^e$Multi-Configuration-Hartree-Fock (MCHF), Saha \citep*{Saha1994};
$^f$S-matrix method, Petrini and de Ara\'{u}jo \citep*{Petrini1994}.
Experimental and theoretical energies (eV),  quantum defects $\mu$ and natural linewidths $\Gamma$ (meV), 
for the two prominent Rydberg series are presented. Lifetimes, $\tau$ (fs), given for the resonance states were 
determined from the uncertainty principle. Bracketed values represent the ALS experimental energies (eV) 
and quantum defects obtained using an average value of $\mu$. 
The error in the calibrated photon energy is $\pm$ 40meV. 
Instrumental resolution of 3800 $\pm$ 150 ($\approx$ 135 $\pm$ 5 meV) was used for the Gaussian 
 portion of the Voigt function when refitting the previous \citep*{Stolte1997} and 
4250 $\pm$ 400  ($\approx$ 124 $\pm$ 12 meV) for the current ALS experimental data.}
\end{deluxetable}

In contrast to this, observed spectra using the LETG  on {\it Chandra},  are calibrated 
to either theoretical calculations or EBIT  measurements \citep*{Schmidt2004,Gu2005}  
for the $1s \rightarrow 2p$ resonance line in H-like, He-like oxygen, Li-like or B-like oxygens ions 
(with all the ensuing complications and difficulties of  ion identification).  
Given the high precision of our experimental and theoretical data we recommend 
that observational data concerning {\it K}-shell photoabsorption of atomic oxygen 
be calibrated to the present work. 

\acknowledgments
BMMcL, and CPB thank the Institute for Theoretical Atomic and Molecular Physics (ITAMP),
 at the Harvard-Smithsonian Center for Astrophysics for their hospitality and 
 support (BMMcL) under the visitor's program.  ITAMP is supported by a grant from the National Science Foundation. 
CPB acknowledges support by US Department of Energy (DoE) grants  through Auburn University. 
WCS acknowledges support from the National Science Foundation under NSF Grant No. PHY-01-40375. 
We thank Professor Alex Dalgarno FRS, Dr John C Raymond, Dr Randall K Smith,
Dr Jeremy J Drake and Dr Brad Wargelin for 
discussions on the  astrophysical applications  and the LETG {\it Chandra} calibration.
Grants of computational time at the National Energy Research Scientific
Computing Center in Oakland, CA, USA,  the  Kraken XT5 facility 
at the National Institute for Computational Science (NICS) in Knoxville, TN, USA
and at the High Performance Computing Center Stuttgart (HLRS) of the University of Stuttgart 
are gratefully acknowledged. The Kraken XT5 facility is a resource of the 
Extreme Science and Engineering Discovery Environment (XSEDE), 
which is supported by National Science Foundation Grant No. OCI-1053575. 
The help of  Hendrik Blum and Tolek Tyliszczak in setting up the experiment 
 on beamline 11 at the ALS is gratefully acknowledged.
The Advanced Light Source in Berkeley, CA, USA, is supported by the Director, 
Office of Science, Office of Basic Energy Sciences, 
of the US Department of Energy under Contract No. DE-AC02-05CH11231.

%\bibliographystyle{apj}                       %% AASTeX
%\bibliography{o-apj}

\begin{thebibliography}{52}
\expandafter\ifx\csname natexlab\endcsname\relax\def\natexlab#1{#1}\fi

\bibitem[{{Angel} \& {Samson}(1980)}]{Samson1988}
{Angel}, G.~C., \& {Samson}, J. A.~R. 1980, \pra,  38, 5578

\bibitem[{{Ballance} \& {Griffin}(2006)}]{ballance06}
{Ballance}, C.~P., \& {Griffin}, D.~C. 2006, {J. Phys. B At. Mol. \& Opt. Phys.},
   39, 3617

\bibitem[{{Ballance et al.}(1999)}]{keith1999}
{Ballance}, C. P.~ {et al.} 1999, \pra,  60, R4217

\bibitem[{{Beiersdorfer et al.}(1999)}]{EBIT1999}
{Beiersdorfer}, P.~ {et al.} 1999, {Rev. Sci. Instru.},   70, 276

\bibitem[{{Berrington et al.}(1995)}]{codes}
{Berrington}, K. A.~ {et al.} 1995, {Comput. Phys. Commun.},  92,
  290

\bibitem[{{Burke}(2011)}]{Burke2011}
{Burke}, P.~G. 2011, {{\it R}-Matrix Theory of Atomic Collisions Application to
  Atomic, Molecular and Optical Processes} (New York, USA: Springer)

\bibitem[{{Clementi} \& {Roetti}(1974)}]{Clementi1974}
{Clementi}, E., \& {Roetti}, C. 1974, {At. Data Nucl. Data Tables},
   14, 177

\bibitem[{{Costantini et al.}(2012)}]{Kaastra2012}
{Costantini}, E.~ {et al.} 2012, \aap,  539,
  A32

\bibitem[{{Fano} \& {Cooper}(1968)}]{Fano1968}
{Fano}, U., \& {Cooper}, J.~W. 1968, Rev. Mod. Phys.,  40, 441

\bibitem[{{Ferland}(2003)}]{Ferland2003}
{Ferland}, G.~J. 2003, \araa,  {41}, 517

\bibitem[{{Fischer}(1991)}]{charlotte1991}
{Fischer}, C.~F. 1991, {Comput. Phys. Commun.},  {64}, 369

\bibitem[{{Foster et al.}(2010)}]{Brickhouse2010}
{Foster}, A. R.~ {et al.} 2010, \ssr,  157, 135

\bibitem[{{Foster et al.}(2012)}]{Foster2012}
---. 2012, \apj,  756, 128

\bibitem[{{Garcia et al.}(2005)}]{Garcia2005}
{Garcia}, J.~ {et al.} 2005, \apjs,  {158}, 68

\bibitem[{{Garcia et al.}(2009)}]{Witthoeft2009}
---. 2009, \apjs,  185, 477

\bibitem[{{Garcia et al.}(2011)}]{Garcia2011}
---. 2011, \apj,  731, L15

\bibitem[{{Gharaibeh et al.}(2011)}]{Soleil2011}
{Gharaibeh}, M. F.~ {et al.} 2011, {J. Phys. B At. Mol. \& Opt. Phys.},
   44, 175208

\bibitem[{{Gorczyca} \& {McLaughlin}(2000)}]{McLaughlin2000}
{Gorczyca}, T.~W., \& {McLaughlin}, B.~M. 2000, {J. Phys. B At. Mol. \& Opt.
  Phys.},  33, L859

\bibitem[{{Gu et al.}(2005)}]{Gu2005}
{Gu}, M. F.~ {et al.} 2005, \apj,  627, 1066

\bibitem[{{Hasoglu et al.}(2010)}]{McLaughlin2010}
{Hasoglu}, M. F.~ {et al.} 2010, \apj,  724, 1296

\bibitem[{{Herman} \& {Skillman}(1963)}]{HS1963}
{Herman}, F., \& {Skillman}, S. 1963, {Atomic Structure Calculations}
  (Englewood Cliffs, NJ, USA: Prentice-Hall)

\bibitem[{{Hinojosa et al.}(2012)}]{Hinojosa2012}
{Hinojosa}, G.~ {et al.} 2012, \pra,
   44, 175208

\bibitem[{{Juett et al.}(2004)}]{Juett2004}
{Juett}, A. M.~ {et al.} 2004, \apj,  612, 308

\bibitem[{{Kallman} \& {Bautista}(2001)}]{Kallman2001}
{Kallman}, T.~R., \& {Bautista}, M.~A. 2001, \apjs,
   134, 139

\bibitem[{{Krause}(1994)}]{Krause1994}
{Krause}, M.~O. 1994, {Nucl. Instr. \& Meth. in Phys. Res. B},  {87},
  178

\bibitem[{{Masuoka} \& {Samson}(1980)}]{Samson1980}
{Masuoka}, T., \& {Samson}, J. A.~R. 1980, \jcp,  77, 623

\bibitem[{{McLaughlin}(2001)}]{McLaughlin2001}
{McLaughlin}, B.~M. 2001, in ASP Con$f$. Series, Vol.  {247},
  {Spectroscopic Challenges of Photoionized Plasma}, ed. G.~J. {Ferland} \&
  D.~W. {Savin} (San Francisco, CA: Astronomical Society of the Pacific), 87

\bibitem[{{McLaughlin} \& {Ballance}(2013)}]{McLaughlin2013}
{McLaughlin}, B.~M., \& {Ballance}, C.~P. 2013, in {McGraw-Hill Yearbook of
  Science and Technology 2013}, ed. {McGraw-Hill} (New York, USA: McGraw-Hill
  Inc), 281

\bibitem[{{McLaughlin} \& {Kirby}(1998)}]{McLaughlin1998}
{McLaughlin}, B.~M., \& {Kirby}, K.~P. 1998, {J. Phys. B At. Mol. \& Opt. Phys.},
   31, 4991

\bibitem[{{Menzel et al.}(1996)}]{Menzel1996}
{Menzel}, A.~ {et al.} 1996, \pra,  54, R991

\bibitem[{{Miyake et al.}(2010)}]{Miyake2010}
{Miyake}, S.  {et al.} 2010, \apj,  709, L168

\bibitem[{{Mueller et al.}(2010)}]{Mueller2010}
{Mueller, A. M.  et al.} 2010, {J. Phys. B At. Mol. \& Opt. Phys.},
 43, 135602

\bibitem[{{Paerels et al.}(2001)}]{Paerels2001}
{Paerels}, F.~ {et al.} 2001, \apj,  546, 338

\bibitem[{{Petrini} \& {de Ara\'ujo}(1994)}]{Petrini1994}
{Petrini}, D., \& {de Ara\'ujo}, F.~X. 1994, \aap,  282, 315

\bibitem[{{Petrini} \& {de Ara\'ujo}(1997)}]{Petrini1997}
{Petrini}, D., \& {de Ara\'ujo}, F.~X. 1997, \aap,
   326, 870

\bibitem[{{Quigley et al.}(1998)}]{keith1998}
{Quigley}, L.~ {et al.} 1998, {Comput. Phys. Commun.}, 114, 225


\bibitem[{{Reilman} \& {Manson}(1979)}]{Reilman1979}
{Reilman}, R.~F., \& {Manson}, S.~T. 1979, \apjs,
   {40}, 85

\bibitem[{{Ressler}(1983)}]{Ressler1998}
{Ressler}, T. 1983, {J. Synchrotron Rad.},  5, 118

\bibitem[{{Robicheaux et al.}(1995)}]{francis95}
{Robicheaux}, F.~ {et al.} 1995, \pra, 52, 1319

\bibitem[{{Saha}(1994)}]{Saha1994}
{Saha}, H.~P. 1994, \pra,  49, 894

\bibitem[{{Sant'Anna et al.}(2011)}]{McLaughlin2011}
{Sant'Anna}, M.~M.~ {et al.} 2011, \prl,  107,  033001

\bibitem[{{Schmidt et al.}(2004)}]{Schmidt2004}
{Schmidt}, M.~ {et al.} 2004, \apj,  604, 562

\bibitem[{{Seaton}(1983)}]{Seaton1983}
{Seaton}, M.~J. 1983, Rep. Prog. Phys.,  46, 167

\bibitem[{{Skinner et al.}(2010)}]{Skinner2010}
{Skinner}, S.~L.~ {et al.} 2010, \aj,  139, 825

\bibitem[{{Stancil et al.}(2010)}]{Stancil2010}
{Stancil}, P. C.~ {et al.} 2010, in {Proceedings of the Dalgarno Celebratory
  Symposium}, ed. {J. F. Babb, K. Kirby and H. Sadeghpour} (London, UK: World
  Scientific Imperial College Press), 102

\bibitem[{{Stolte et al.}(1997)}]{Stolte1997}
{Stolte}, W. C.~ {et al.} 1997, {J. Phys. B At. Mol. \& Opt. Phys.},
   30, 4489

\bibitem[{{Stolte et al.}(2008)}]{Stolte2008}
---. 2008, {J. Phys. B At. Mol. \& Opt. Phys.},  41, 145102

\bibitem[{{Verner et al.}(1993)}]{Verner1993}
{Verner}, D.~A.~ {et al.} 1993, {At. Data Nucl. Data Tables},  55,  233

\bibitem[{{Wiley} \& {McLaren}(1955)}]{Wiley1955}
{Wiley}, W.~C., \& {McLaren}, I.~H.~1955, {Rev. Sci. Inst.},  26, 1050

\bibitem[{{Yeh}(1993)}]{Yeh1993}
{Yeh}, J. 1993, {Atomic Calculation of Photoionization Cross-Sections and Asymmetry Parameters} (New York, USA: Gordon \& Breach )
\end{thebibliography}

\end{document}